\documentclass{pasa}
\title[MWA GW Searches]{Strategies  for Finding Prompt Radio Counterparts to
  Gravitational Wave Transients with the Murchison Widefield Array}
\author[Kaplan et al.]{D.~L.\ Kaplan$^1$, T.~Murphy$^{2,3}$,
 A.~Rowlinson$^{4,5}$, S.~D.~Croft$^{6,7}$,  R.~B.~Wayth$^{3,8}$, and C.~M.~Trott$^{3,8}$\\
  \affil{$^1$Department of Physics, University of
  Wisconsin--Milwaukee, Milwaukee, WI 53201, USA}%
\affil{$^2$Sydney Institute for Astronomy, School of Physics, The University of Sydney, NSW 2006, Australia}
\affil{$^3$ARC Centre of Excellence for All-sky
  Astrophysics (CAASTRO)}
\affil{$^4$Anton Pannekoek Institute for Astronomy,
  University of Amsterdam, Science Park 904, 1098 XH Amsterdam, The
  Netherlands}
\affil{$^5$ASTRON, The Netherlands Institute for Radio Astronomy, Postbus 2, 7990 AA, Dwingeloo, The Netherlands}
\affil{$^6$University of California, Berkeley, Astronomy Dept.,
501 Campbell Hall \#3411, Berkeley, CA 94720, USA}
\affil{$^7$Eureka Scientific, Inc., 2452 Delmer Street Suite 100, Oakland, CA 94602, USA}
\affil{$^8$International Centre for Radio Astronomy Research, Curtin University,
Bentley, WA 6102, Australia}
}
\jid{PASA}
\jyear{\the\year}

\newcommand{\plotone}[1]{\includegraphics[width=\columnwidth]{#1}}
\def\degr {\ensuremath{^{\circ}}}
\newcommand{\I}{\ensuremath{I_{\rm MWA}}}

\usepackage[authoryear]{natbib}
\bibpunct{(}{)}{;}{a}{}{,}
\usepackage{aas_macros}

\begin{document}

\begin{abstract}
  We present and evaluate several strategies to search for prompt, low-frequency radio
  emission associated with gravitational wave transients using the
  Murchison Widefield Array (MWA).  As we are able to repoint the MWA
  on timescales of tens of seconds, we can search for the dispersed
  radio signal that has been predicted to originate along with or
  shortly after a neutron star-neutron star merger.  We find that
  given the large, $600\,{\rm deg}^2$ instantaneous field-of-view of
  the MWA we can cover a significant fraction of the predicted
  gravitational wave error region, although due to the complicated
  geometry of the latter we only cover $>50$\% of the error region for
  approximately 5\% of events, and roughly 15\% of events will be
  located $<10\degr$ from the MWA pointing center such that they will
  be covered in the radio images.  For optimal conditions our limiting
  flux density for a 10-s long transient would be 0.1\,Jy, increasing
  to about 1\,Jy for a wider range of events.  This corresponds to
  luminosity limits of $10^{38-39}\,{\rm erg\,s}^{-1}$ based on
  expectations for the distances of the gravitational wave transients,
  which should be sufficient to detect or significantly constrain a
  range of models for prompt emission.
\end{abstract}

\begin{keywords}
gravitational waves -- gamma-ray burst:general -- methods:
observational -- radio continuum: general 
\end{keywords}

\maketitle%

\section{Introduction}
In 2015 September, the LIGO/Virgo Consortium (LVC) began its O1
science run that resulted in the first detection of gravitational
waves \citep[][also see \citealt{ligo16b} for a second event]{ligo16}.  Together with the gravitational wave (GW)
analysis, the LVC sent private alerts to the electromagnetic (EM)
followup community \citep{em2016,em2016b} to identify coincident
electromagnetic transients (e.g.,
\citealt{2016arXiv160501607L,2016MNRAS.460L..40E,2016ApJ...822L...8T,2016ApJ...823L...2A,2016arXiv160503216M,2016ApJ...820L..36S,2016arXiv160203920C}).
This identification is complicated by the very large uncertainty
regions for the GW events: as discussed in \citet{2014ApJ...795..105S}
and \citet{2014ApJ...789L...5K}, the error regions for these events
can cover hundreds of square degrees (especially for the initial
sensitivities of the detectors).  Moreover, they need not be compact
or simply connected.  While identifying an EM counterpart would
greatly enhance the utility of the GW signal
\citep[e.g.,][]{2009astro2010S.235P,2014ApJ...795..105S,2016MNRAS.459..121C,2016arXiv160408530B}
and would enable a range of new physical and astrophysical tests, it
is not a simple task
\citep[e.g.,][]{2014ApJ...789L...5K,2015ApJ...814...25C}.

The EM counterparts span a range of models at a range of wavelengths:
see \citet{2012ApJ...746...48M}, \citet{2014ApJ...795..105S},
\citet{2014ApJ...789L...5K}, and \citet{2016MNRAS.459..121C}, among
other recent publications.  At low radio frequencies telescopes such
as the Murchison Widefield Array (MWA; \citealt{tin13}), the Low
Frequency Array (LOFAR; \citealt{2013A&A...556A...2V}), and the Long
Wavelength Array (LWA; \citealt{2009IEEEP..97.1421E}) have a number of
advantages over optical/infrared searches: they have fields-of-view of
hundreds to thousands of square degrees; unlike the
optical/near-infrared sky which has a large number of transients
present in every field \citep[e.g.,][]{2015ApJ...814...25C}, the radio
sky is relatively quiet at these frequencies
\citep{2015MNRAS.452.1254K,2015AJ....150..199T,2016MNRAS.456.2321S,rowlinson16,2016arXiv160400667P}
with very few unrelated transients
\citep[e.g.,][]{2016arXiv160509395H} to distract from those associated
with the GW event; and  many of the low-frequency facilities
have no moving elements and so in principle can respond within seconds
to an external trigger.

While most expectations for transients associated with GW sources
at radio wavelengths have concentrated on late-time radio afterglows
and remnants
\citep{2015ApJ...806..224M,2015MNRAS.450.1430H,2016arXiv160205529M,2016arXiv160509395H,2016arXiv160806518P},
which only peak after hundreds or thousands of days at 150\,MHz and
can be quite faint (depending on the parameters of the explosion and
the circum-burst medium), there are models that
predict a prompt, coherent radio transient from the GW event
\citep[e.g.,][]{1996A&A...312..937L,2000AA...364..655U,2010ApSS.330...13P,2013PASJ...65L..12T,2014ApJ...780L..21Z,2016ApJ...822L...7W,2016MNRAS.461.4435M}
which may be related to the phenomenon of fast radio bursts (FRBs;
\citealt{2007Sci...318..777L,2013Sci...341...53T}) at least some of
which may be cosmological in origin \citep[][although see
  \citealt{2016ApJ...821L..22W,2016arXiv160304421V}]{2016arXiv160207477K}.
Searches for direct connections between GW events and FRBs are
proceeding largely\footnote{Even without a direct connection, current
  and future population studies
  \citep{2015MNRAS.447.2852K,2015ApJ...807...16L,2016MNRAS.455.2207R,2016arXiv160206099L}
  may be able to argue statistically for or against a connection
  \citep[e.g.,][]{2013Sci...341...53T,2016arXiv160204542Z,2016ApJ...825L..12C}.}
through searches for GW events associated with individual FRBs
\citep[e.g.,][]{2016arXiv160501707T} since the GW detectors have
quasi-all sky sensitivity.  But this strategy can be reversed: given
their wide fields of view and very fast response times
\citep{2015ApJ...814L..25K}, low-frequency facilities might be ideal
for finding such prompt emission
\citep{2016MNRAS.459..121C,2015PASA...32...46H} triggered instead by
the GW signal.  We then must optimize the  followup procedure
to maximize the prospects of a discovery without time for human-aided
decision making.

Strategies to aid followup have been studied in the
optical/near-infrared regime
\citep{2016arXiv160301689R,2015arXiv150604035C} where the signals are
likely to be faint, relatively short in duration, and may be quite red
\citep{2010MNRAS.406.2650M,2012ApJ...746...48M,2013ApJ...775...18B,2014MNRAS.441.3444M,2015MNRAS.450.1777K}.
In the optical/near-IR the search can be aided by using prior
information on host galaxies and likely distances to help reduce the
search volume
\citep[e.g.,][]{2016ApJ...820..136G,2016arXiv160307333S}.  Strategies
have also been studied in the X-ray regime
\citep[e.g.,][]{2016MNRAS.455.1522E}, looking to directly probe the
association between GW events and short gamma-ray bursts.

In this paper we present and compare concrete strategies for
low-frequency radio followup to search for prompt radio emission from
a GW transient, where we use the MWA as an example to determine the
likely sensitivity and success rate.  Unlike in the
optical/near-infrared, where a limited time window nonetheless allows
limited sky coverage \citep[e.g.,][]{2015arXiv150604035C}, if we are
searching for a prompt (duration $\lesssim$ms) radio signal we are
limited to only a single pointing, and so we must optimize where that
is with limited information.  We discuss this in the context of
simulated GW error regions from the first couple of years of GW
observations, using two and three detectors (based on the simulated
events from \citealt{2014ApJ...795..105S}).  The MWA occupies a middle
ground in the current generation of low-frequency arrays: it has a
considerably wider field of view than the more sensitive LOFAR, but it
is pointed, unlike the LWA which observes the whole (visible) sky.
The MWA can respond on timescales of seconds to external triggers,
which is currently not possible with LOFAR (A.~Rowlinson 2016,
pers.~comm.) and which is not needed with the LWA's all-sky coverage.
We discuss a search that focuses on standard imaging techniques
\citep{2015AJ....150..199T,2015ApJ...814L..25K,rowlinson16} and not
more rapid ``beam-formed'' data
\citep[e.g.,][]{2014A&A...570A..60C,2015MNRAS.452.1254K,2015PASA...32....5T},
which, although it can be more sensitive to fast signals, is much more
computationally intensive to process.
  
\section{Search Metrics}
\label{sec:search}
To assess the success or failure of our pointing strategies, we looked at
a number of different metrics and computed these for simulated GW
events based on realistic expectations for the first two years of GW
detector operations \citep{2014ApJ...795..105S}\footnote{Given at:
  \texttt{http://www.ligo.org/scientists/first2years/}.}.
These simulated events included the large uncertainty region that will
be communicated rapidly to EM observers, as well as the actual event
locations and distances.  Therefore, for a given pointing strategy, we
 computed the distribution of flux density and luminosity
sensitivities for each simulated event.  These were compared
with model predictions \citep[such
  as][]{2010ApSS.330...13P,2013PASJ...65L..12T,2014ApJ...780L..21Z}.
We  also computed the separation between our pointing center and
the event location or the fraction of the total probability map
covered by the observations \citep{em2016,em2016b}, but these are of
limited use for a wide-field aperture array.  This is because unlike
optical observations with a finite but relatively uniformly covered
field-of-view (limited slightly by vignetting), the sensitivity for an
aperture array over the sky is controlled primarily by the primary
beam and varies considerably over the area imaged, with some
sensitivity even down to the horizon
\citep[e.g.,][]{2015RaSc...50..614N}.

The flux density sensitivity of the MWA observations at the positions
of the GW events was computed using  the tile area at 150\,MHz from \citet{tin13},
along with a receiver temperature of 50\,K. To that we added a
predicted sky temperature, computed by integrating the Global Sky
Model \citep[][interpolated to 150\,MHz]{2008MNRAS.388..247D} over the
MWA tile response.  We do not account for the additional contribution
of the Sun.  We assume a 10\,$\sigma$ detection threshold with 30\,MHz
bandwidth over a 10\,s observation.  The search duration is determined
from likely dispersion measures: for an event at a redshift $z\approx
0.05$ which is a typical horizon for the current detectors,
\citep{2003ApJ...598L..79I,2004MNRAS.348..999I} predict an
extragalactic dispersion
measure DM$\approx 50\,{\rm pc\,cm}^{-3}$.  This can be added to
dispersion measures of $50-100\,{\rm pc\,cm}^{-3}$ from the Milky Way
and the host galaxy
\citep[e.g.,][]{2002astro.ph..7156C,2016arXiv160207477K,2015ApJ...814L..25K} for a total
of $150-250\,{\rm pc\,cm}^{-3}$.  With the 30\,MHz bandwidth of the
MWA centered at 150\,MHz, this gives an event duration of 10--20\,s
\citep{2004hpa..book.....L,2015ApJ...814L..25K,rowlinson16}. We assume no
additional loss of sensitivity due to the complex nature of the
Galactic synchrotron emission, but this is likely reasonable given the
short duration of the expected signals.  The resulting flux density is
converted to a luminosity using the simulated event distance.

\begin{figure*}
  \centerline{
    \includegraphics[width=0.5\textwidth]{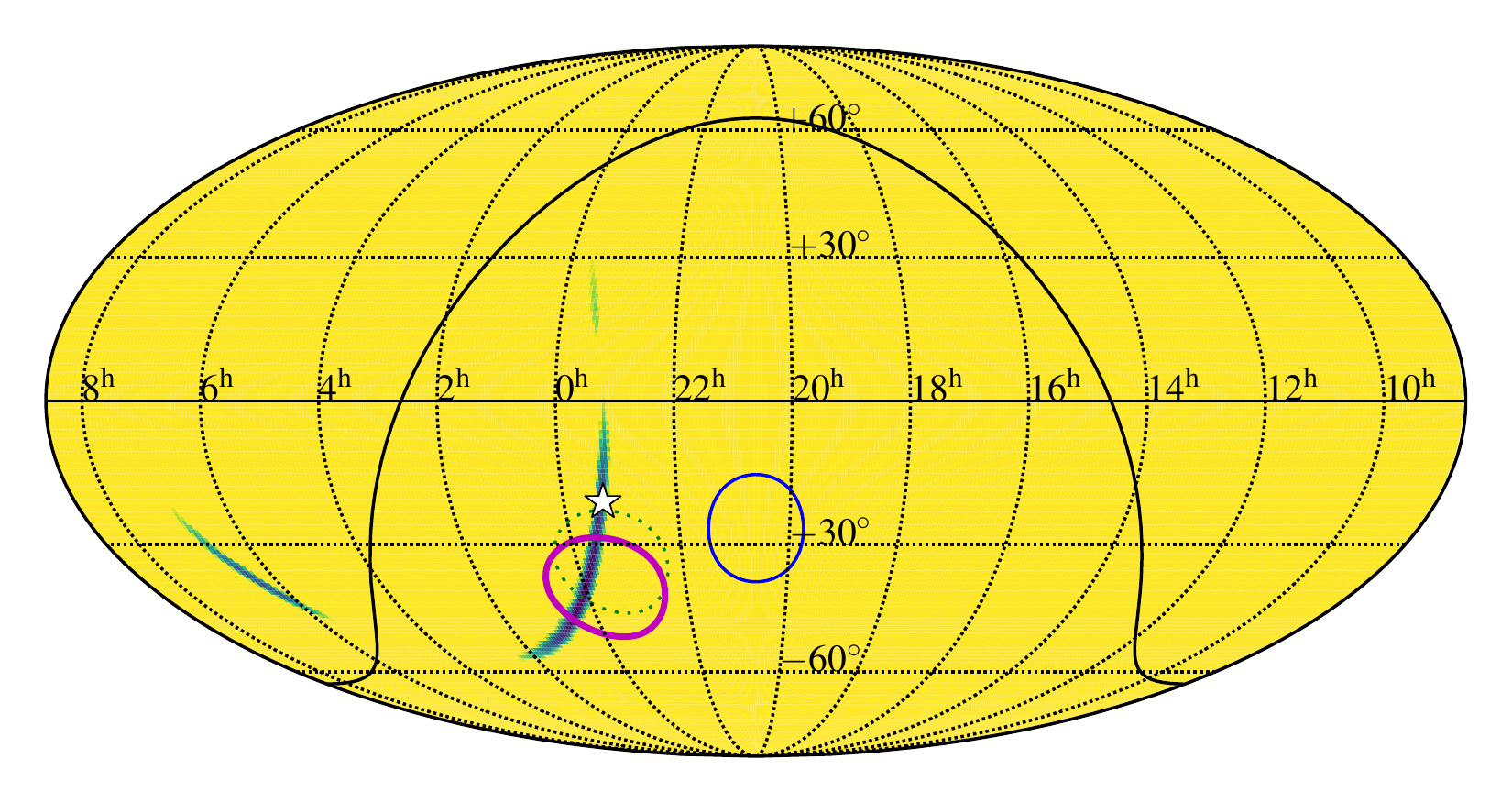}%
    \includegraphics[width=0.5\textwidth]{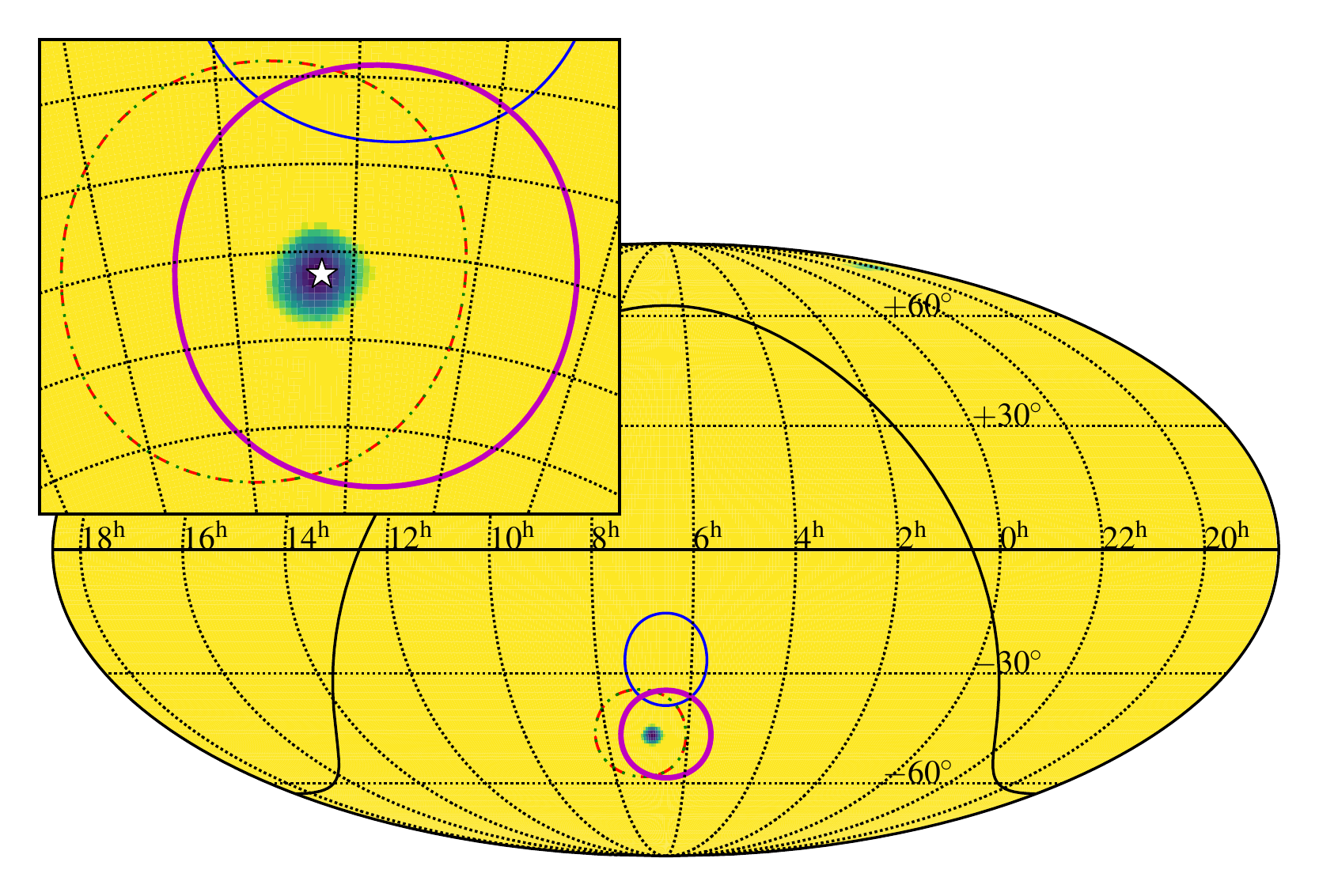}
  }
  \caption{Sky probability map of simulated LVC transients from
    \citet{2014ApJ...795..105S}.  The color is proportional to the
    $\log_{10}$ of the probability.  The black lines are the MWA
    horizon.  The MWA half-power beams are shown by the blue lines
    (strategy 1: zenith), red dashed lines (strategy 2: maximum
    probability), green dotted lines (strategy 3: maximum probability
    weighted by $\cos^2 Z$), and thick magenta lines (strategy 4:
    maximum \I). The white stars are the actual event locations.  For
    the event on the left the GW signal was only recovered by two
    detectors with a net signal-to-noise ratio of 14.7, leading to a
    large error region.  In contrast the event on the right the GW
    signal was recovered by three detectors with a net signal-to-noise
    ratio of 21.8, which greatly improves the localizations and leads
    to very similar  pointings for strategies 2--4.
    The
    images are Mollweide projections of the celestial sphere, labeled
    in Right Ascension and Declination, and centered on the local
    sidereal time at the MWA.  For the event on the right we also show
    a zoom around the position of the event.}
  \label{fig:events}
\end{figure*}

\section{Pointing Strategies}
To point the MWA we change the delays for individual dipole antennas
on each tile.  The whole array (128 tiles) can be pointed together, or
it can be split into subarrays, but only a single pointing is current
possible for each tile.  The pointing is generally done at a series of
discrete steps about $7\degr$ apart.
The nominal field-of-view is about $600\,{\rm
  deg}^2$ at 150\,MHz \citep{tin13}.  Our first goal was to determine
a pointing strategy: when a \texttt{HEALPIX}
\citep{2005ApJ...622..759G} sky probability map is received, where do
we point the MWA and do we point as a single array or use subarrays?  

We consider several simple strategies:
\begin{enumerate}
\item Zenith pointing
\item Point toward the maximum of the probability accessible (i.e., above the
  horizon) in the map
\item Point toward the maximum of the  probability weighted by $\cos^2Z$ accessible (i.e., over the
  horizon) in the map, where $Z$ is the zenith angle
\item Maximize the overlap between the MWA primary beam pattern and
  the GW probability map
\end{enumerate}
The first of these serves as a benchmark.  In addition, the
sensitivity of the MWA is maximum at zenith and the primary beam at
that pointing has been characterized considerably better than for
other pointings.  Finally, this strategy has the benefit that no
decisions are necessary, so the MWA can repoint as soon as a GW alert
is received.  Moreover, meridian drift-scans are among the most common
observational program (e.g., for the MWA Transients Survey and the
GaLactic and Extragalactic All-Sky MWA Survey;
\citealt{2016MNRAS.461..908B,2015PASA...32...25W}) so if we did not
interrupt an ongoing observation this would be the most likely result.

The second strategy is also simple.  We simply identify the point in
the GW probability map (which is sent along with the alert
announcements) that has the highest value and which is also
above the horizon.  This is also relatively fast to compute, although
it does require parsing of the GW probability map and not just
knowledge of an alert.

The third strategy is very similar to the second, except that we
account for the overall envelope of a Hertzian dipole which is the
basic component of an MWA tile
\citep{tin13,2015RaSc...50...52S,2015RaSc...50..614N}.  This
downweights observations close to the horizon.

Finally, the fourth strategy examines the overlap between the LVC GW
sky probability and the MWA's pointing pattern.  Specifically, it
tries to maximize:
\begin{equation}
\I=  \int d\Omega\, P_{{\rm LVC}}(\alpha,\delta) B_{\rm
      tile}(\alpha,\delta)
\label{eqn:overlap}
\end{equation}
where $P_{\rm LVC}$ is the normalized sky probability returned in the
LVC \texttt{HEALPIX} map as a function of sky position
$(\alpha,\delta)$, and $B_{\rm tile}(\alpha,\delta)$ is the individual
tile response for the MWA, normalized to 1 at the zenith.
Constructed in this way, we  maximize \I\ by 
choosing the best discrete pointing $B_{\rm tile}$.

The implementation of the four strategies proceeds as follows.
Strategy 1 is fixed to the zenith, so no computation is necessary.
For the other strategies, we first compute the MWA horizon.  If the
integral of $P_{\rm LVC}$ (weighted by $\cos^2Z$ for strategy 3, or by
$B_{\rm tile}$ for strategy 4) above the horizon is less than some
threshold (currently 2\%) we do not consider the target worthwhile,
and do not return a pointing position.  Otherwise, strategies 2 and 3
return the discretized pointing closest to the maximum of $P_{\rm
  LVC}$ (strategy 2) or $P_{\rm LVC}\cos^2Z$ (strategy 3).  For
strategy 4, we iterate through the
range of discrete tile pointings.  For each one we compute the normalized MWA
tile beam pattern sampled on the \texttt{HEALPIX} grid\footnote{For speed, we
can resample the LVC \texttt{HEALPIX} grid from the initially fine
resolution down to a coarser resolution suitable for the MWA.  For
example, often the LVC maps are returned with \texttt{NSIDE}$=2048$,
corresponding to a pixel size of $0.029\degr$.  This is often smaller
than a single MWA pixel, and with $50\times 10^6$\,pixels, the
calculation can be slow.  Instead we resample (conserving probability)
down to \texttt{NSIDE}$=64$, corresponding to a pixel size of
$0.92\degr$ (and 49,152\,pixels)}.  We then identify the pointing
position that maximizes the integral of $P_{\rm LVC}\times B$.  If
that integral is less than a threshold (again currently 2\%) we again
do not consider the target worthwhile, and do not return a pointing
position.  Otherwise we return the optimal target position
$\alpha,\delta$, along with (if desired) the beamformer delays and the
integrated probability.

\section{Event Expectations}
We sought to predict our event coverage and flux density/luminosity
sensitivites to actual LIGO transients using the simulated events from
2016 (including potentially both LIGO sites as well as Virgo) 
\citep{2014ApJ...795..105S}.  There are 475 simulated signals,
covering a realistic range of signal-to-noise ratio and sky position.
Examples of this are shown in Figure.~\ref{fig:events}.  It is clear
that even with a reasonable probability coverage it is possible to
miss the actual transient.  We see two qualitatively different
results.  In the first, especially where only two GW detectors see the
transient, the large, elongated uncertainty region means that there is
a substantial chance of missing the actual transient regardless of
strategy, even with the MWA's large field-of-view, but on average
strategies 2--4 will see some fraction of the transients discussed below. The second type of
result has a small enough uncertainty region that all pointing
strategies give substantially similar results, and we are just limited
by the horizon and the sensitivity of the MWA.

\begin{figure}
  \plotone{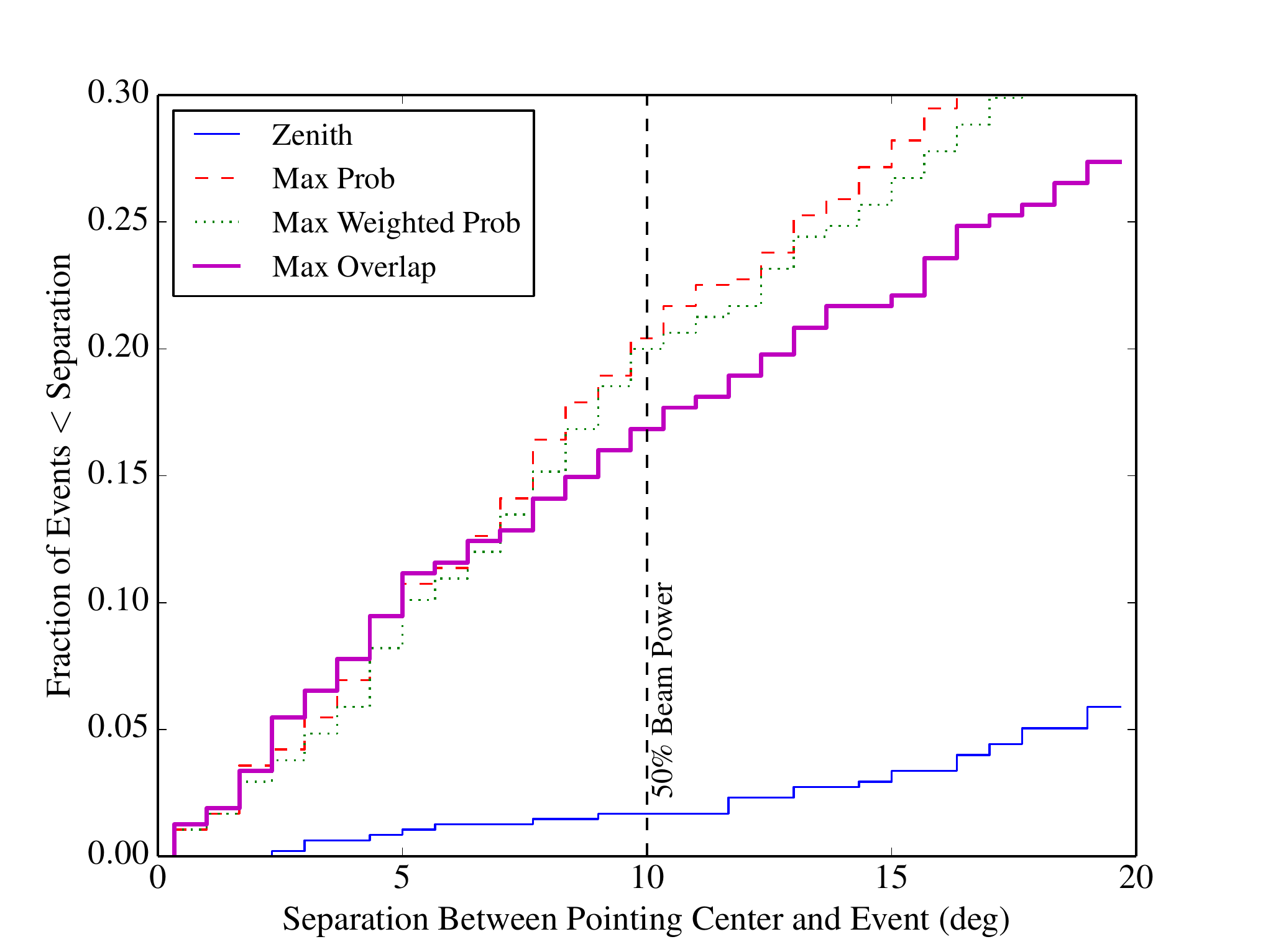}
  \caption{Cumulative histogram of $\theta$ for the simulated
    2016 events, assuming observations at 150\,MHz: blue lines
    (strategy 1: zenith), red dashed lines (strategy 2: maximum
    probability), green dotted lines (strategy 3: maximum probability
    weighted by $\cos^2 Z$), and thick magenta lines (strategy 4:
    maximum \I).
    The vertical line
  is the half-power point for 150\,MHz.}
  \label{fig:sep}
  \end{figure}

\begin{figure}
  \plotone{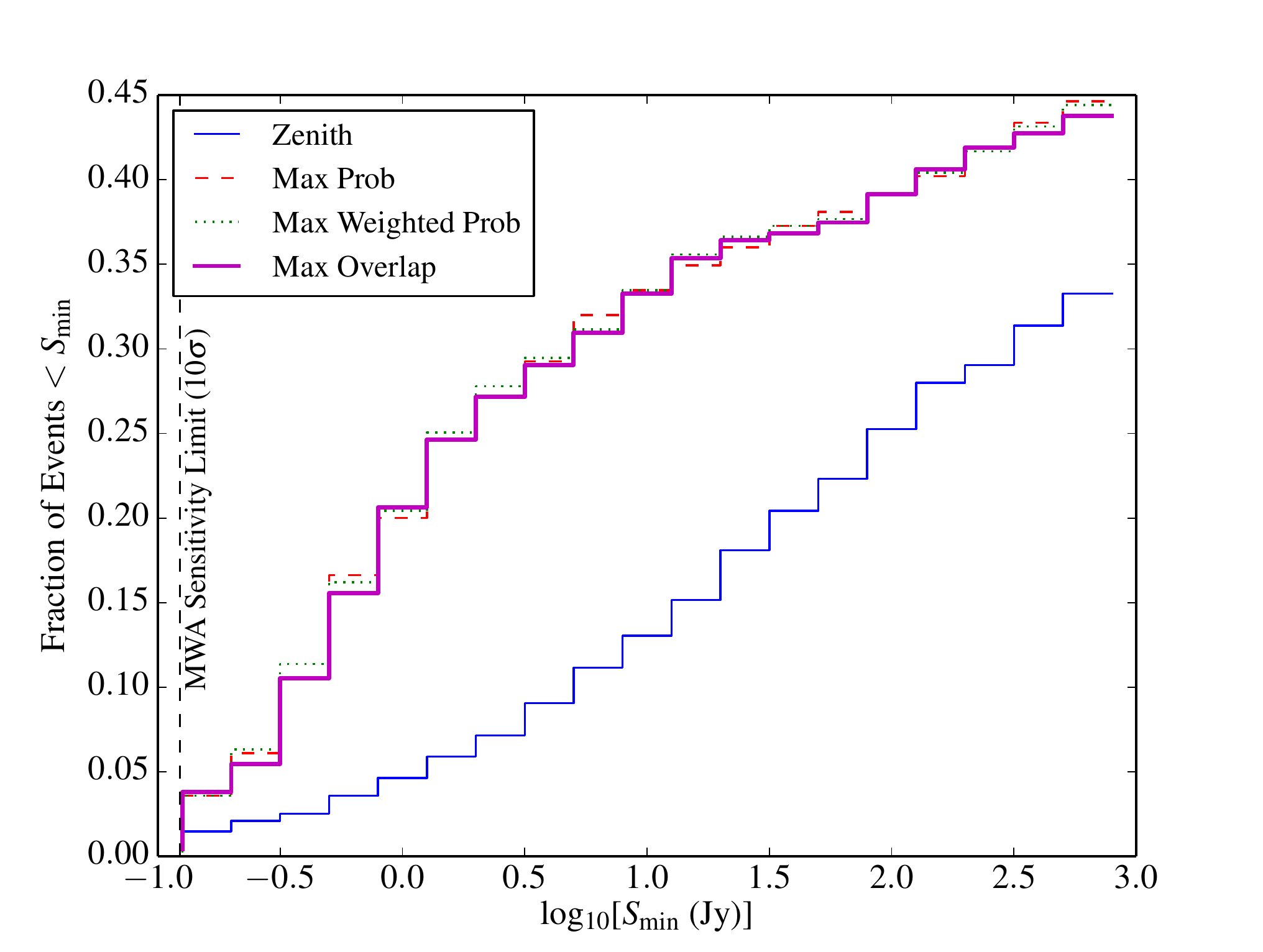}
  \caption{Cumulative histogram of limiting flux density (in Jy) for the simulated
    2016 events, assuming observations at 150\,MHz: blue lines
    (strategy 1: zenith), red dashed lines (strategy 2: maximum
    probability), green dotted lines (strategy 3: maximum probability
    weighted by $\cos^2 Z$), and thick magenta lines (strategy 4:
    maximum \I). This assumes a
    10\,$\sigma$ detection over 30\,MHz of bandwidth in a 10\,s
    integration.  The sky temperature has been computed by integrating
    the Global Sky Model \citep[][interpolated to
      150\,MHz]{2008MNRAS.388..247D} over the MWA tile response and
    assumes an additional 50\,K for the receiver temperature.  
    The vertical line
  shows the nominal 10\,$\sigma$ sensitivity limit from \citet{tin13}.}
  \label{fig:flux}
  \end{figure}

\begin{figure}
  \plotone{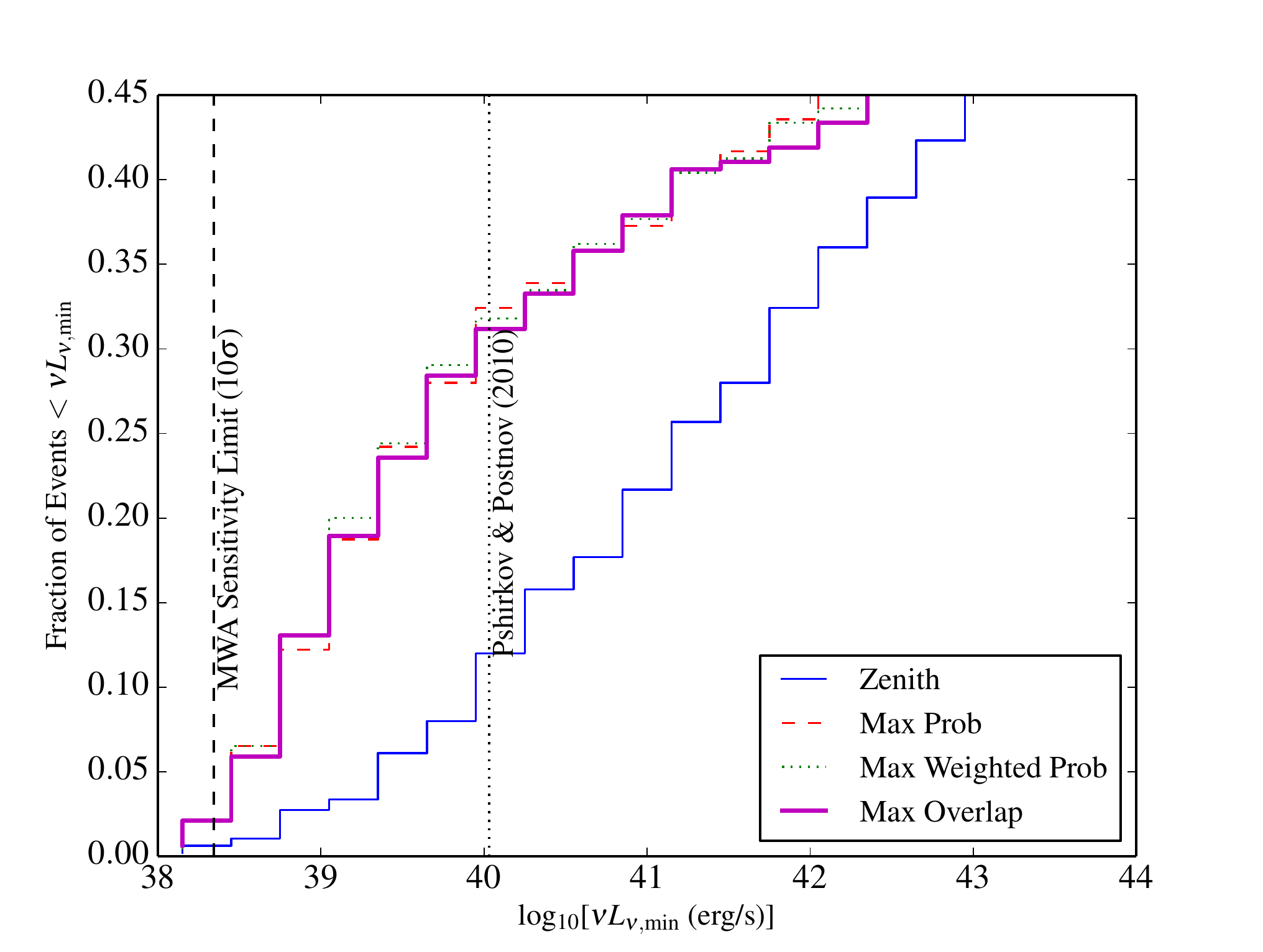}
  \caption{Cumulative histogram of limiting luminosity $\nu L_\nu$ (in
    erg/s) for the simulated 2016 events, assuming observations at
    150\,MHz: blue lines
    (strategy 1: zenith), red dashed lines (strategy 2: maximum
    probability), green dotted lines (strategy 3: maximum probability
    weighted by $\cos^2 Z$), and thick magenta lines (strategy 4:
    maximum \I).  This assumes a 10\,$\sigma$ detection over 30\,MHz of
    bandwidth in a 10\,s integration.  The sky temperature has been
    computed by integrating the Global Sky Model \citep[][interpolated
      to 150\,MHz]{2008MNRAS.388..247D} over the MWA tile response and
    assumes an additional 50\,K for the receiver temperature.  The
    dashed vertical line shows the nominal 10\,$\sigma$ sensitivity limit from
    \citet{tin13} at a distance of 100\,Mpc, while the dotted vertical
  line shows the predicted luminosity from \citet{2010ApSS.330...13P}.}
  \label{fig:luminosity}
  \end{figure}

We show the results in Figures~\ref{fig:sep}--\ref{fig:luminosity}.  
In  Figure~\ref{fig:sep} we plot the separation between the MWA
pointing center for each event and each strategy with the actual event location.
For strategies 2 and 3, roughly 20\% of the events have
separations $<10\degr$, which is the half-power point at 150\,MHz.
This decreases to roughly 15\% of the events for strategy 4, and
$<3$\% for strategy 1 (the control).  In many cases the pointings will
be similar for strategies 2--4 (as in the right panel of
Figure~\ref{fig:events}), which accounts for the very similar
distribution of events at separations $<5\degr$.  

Using these results we can also compute the expected flux density
sensitivity of the MWA observations at the positions of the GW events,
as described in \S~\ref{sec:search}.  For the coldest parts of the
sky away from the Galactic plane, we get a limiting flux density of
about 0.1\,Jy.  However, given the influence of Galactic synchrotron
emission and the limited collecting area away from zenith, only 5\% of
the simulated events are close to that limit.  If we consider the 15\%
of events that are with the half-power point, a more typical limiting
flux density is 1\,Jy (Fig.~\ref{fig:flux}).  This can be compared
with predictions from e.g., \citet{2010ApSS.330...13P}, who claim
something like $S\approx 6\,$Jy at a distance of 100\,Mpc and a
frequency of 150\,MHz, so we should be able to see events like those
in about 15\% of the cases.  Once again we see little difference
between strategies 2--4.

\begin{figure*}
  \includegraphics[width=\textwidth]{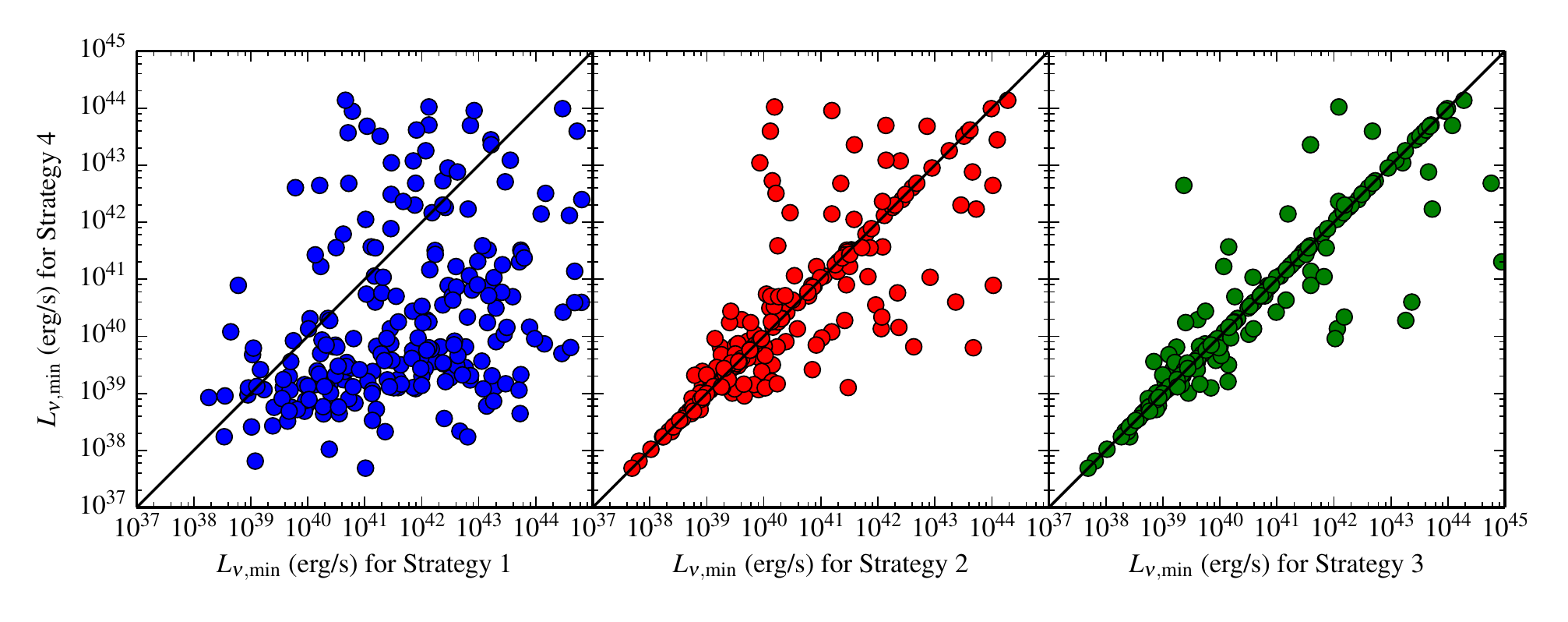}
  \caption{Comparison of limiting luminosity $\nu L_\nu$ (in
    erg/s) for the simulated 2016 events for each strategy, assuming observations at
    150\,MHz: left
    (strategy 1: zenith), middle (strategy 2: maximum
    probability), right (strategy 3: maximum probability
    weighted by $\cos^2 Z$), all compared to strategy 4.  }
  \label{fig:comparison}
  \end{figure*}

We can also compute the limiting luminosity using the simulated GW
event distances, finding limits of $10^{38-39}\,{\rm erg/s}$
(Fig.~\ref{fig:luminosity}).  As a representative comparison, we use
the signal predicted by \citet[]{2010ApSS.330...13P}: 6\,Jy at
100\,Mpc, at a frequency of 150\,MHz (this assumes an intrinsic
spin-down luminosity of $\dot E=10^{50}\,{\rm erg\,s}^{-1}$,
efficiency scaling exponent $\gamma=0$, and that the burst is
scattered to a duration of $\sim 10\,$s).
For roughly 30\% of the events would
we be able to see the such a signal. This agrees with the estimates
presented in \citet{2015ApJ...814L..25K}, where the prompt emission
predicted by various models \citep[such
  as][]{2010ApSS.330...13P,2013PASJ...65L..12T,2014ApJ...780L..21Z} is
compared against the sensitivity of the MWA for prompt searches.
Overall they find that the sensitivity of the MWA should be sufficient
for events at a redshift of $z=0.05$, given the available predictions.
In Figure~\ref{fig:comparison} we show the event-by-event comparison
for the different strategies.  Not unexpectedly, strategy 1 performs
poorly. Comparing strategies 2 and 3 to 4, the spread is a lot smaller
for the better events (those with luminosity limits $\lesssim
10^{39}\,{\rm erg\,s}^{-1}$), since those tend to have smaller
uncertainty regions that are well covered by all three strategies.
For the remaining events the results change significantly whether
strategy 2/3 or 4 is used, but there is not a global preference for
one or the other.

We explored the frequency dependence of these limits by repeating the
exercise above for observations at 120\,MHz, 150\,MHz, and 180\,MHz,
which are the range where the MWA's sensitivity is optimized.  At
lower frequencies our primary beam will be larger and we will cover
more of the GW error region.  Conversely, at higher frequencies the
sky noise is lower, so the same observation will reach a lower
limiting flux density.  The 20$^{\rm th}$ percentile for the limiting
flux density (corresponding to a nominal 1\,Jy in Fig.~\ref{fig:flux})
is about 40\% higher at 120\,MHz compared to 150\,MHz, and 20\% higher
at 150\,MHz compared to 180\,MHz.  However, we must also correct for
the intrinsic spectral index $\beta$ (with $S_\nu\propto \nu^\beta$),
which is predicted to vary between $-1$ and $-2$ (see e.g.,
\citealt{2015ApJ...814L..25K}).  If this spectral index is steeper
than $-1.5$ then the lower frequencies will dominate.  This leads us
to a  preference for lower-frequency observations, but the unknown
spectral index makes this preference weak.

\section{Discussion \& Conclusions}
\label{sec:discuss}

In the analysis above, we found  that about 15\%--20\% of the
events would be within the MWA's half-power point.  Therefore, we
would require follow-up of 
$\gtrsim 6$ events before we have one with a relatively sensitive
observation down to luminosity limits of $\sim 10^{39}\,{\rm
  erg\,s}^{-1}$.  Currently the predicted rates of neutron
star-neutron star inspirals are 0.4--400\,${\rm yr}^{-1}$ with a mean
of 40\,${\rm yr}^{-1}$
\citep{2010CQGra..27q3001A,2015ApJ...806..263D,2016LRR....19....1A},
so a single year of observing should be sufficient for one or more
constraining observations if the rates are not too close to the most
pessimistic case.

Comparing the four strategies outlined in \S~\ref{sec:search}, we can
easily reject the control strategy 1 (zenith pointing), but the
remaining strategies are largely comparable in performance.
Individual events may be seen better with one or the other but with
the limited information available from the prompt GW triggers we
cannot know which will be best in advance.

Regardless of which strategy, the metrics in \S~\ref{sec:search} rely
on the MWA being sensitivity-limited rather than event-limited.  In
principle we could have different tiles with different pointing
locations, so as to cover the large LVC uncertainty region.  However,
the limited collecting area of the MWA drives us to point all of the
tiles in a single subarray so as to achieve the most sensitive
possible observation, rather than attempt to cover more of the GW
error region at lower sensitivity \citep[cf.][]{2015ASPC..496..254B}.
This is because, unlike in the optical regime where prompt emission
from gamma-ray bursts is a known (albeit rare) phenomenon
\citep[e.g.,][]{2014Sci...343...38V}, with a known luminosity
function, prompt radio emission from a gamma-ray burst or a GW event
has never been seen \citep{2012ApJ...757...38B,2015ApJ...814L..25K} so we do not know if
shallower observations will be adequate: in the best 30\% of the cases where
the MWA did cover the GW event with a reasonable sensitivity, our
luminosity limits were only $\sim 1$ order of magnitude below model
predictions.  Splitting the MWA into subarrays would mean that all
observations were less constraining. We can instead make up for
the possibility of missing the GW event in a statistical sense by
observing a larger number of events.  
At the same time the increasing performance of the GW detectors will
lead to a large number of targets with improving localization.
Therefore we believe it best to stick with a single array, but this
will be re-evaluated as the actual successes are evaluated.
Similarly, we could experiment with other observational modes like
splitting our 30\,MHz bandpass into multiple sub-bands (as in
\citealt{2015ApJ...814L..25K}), which could be advantageous if a
bright but frequency-dependent signal is expected.  Given the degree
of uncertainty about these models that is unlikely to be preferred at
least to start, but as we gain experience we may change our procedure.

We implemented  strategy 4 for the MWA during the first LIGO
observing run (O1; 2015 September to 2016 January) covering the first
detection, GW~150914.  However, this trigger was released after a
considerable delay (several days) needed for human examination of the
event.  Therefore we did not require any real-time decisions about
strategy but instead could use multiple pointings to tile the GW error
regions \citep{em2016}.  We expect that as the LVC improves their
internal vetting and pipelines their latency will improve to
90--120\,s after the GW event \citep{2014ApJ...795..105S,T1400054-v7}
or possibly better \citep{2012ApJ...748..136C,2016MNRAS.459..121C} and
this strategy will be employed.

It is worth noting that the first published GW signal is from a binary
black hole system \citep{ligo16}, which is not expected to have any EM
signature \citep[][but see \citealt{2016arXiv160203920C}]{em2016}.
The rates of similar events will likely be quite high once LIGO
reaches its full design sensitivity, approaching 1/day.  If this is
the case then we will certainly have the opportunity to cover a
sufficient number of error regions to search statistically for any
associated EM emission, although the greater distances to the more
massive systems will limit our sensitivity.

As discussed in \citet{2015ApJ...814L..25K} and
\citet{2016MNRAS.459..121C}, the expected delay of the radio signal
relative to the GW transient is tens of seconds up to several minutes,
based on the simulated distances of the transients and the expected
extragalactic plus galactic dispersion measures.  The actual time of
any prompt radio signal may also be shifted by up to tens of seconds
\citep[e.g.,][]{2014ApJ...780L..21Z}, potentially in either direction.
Given the fast, $\approx 16\,$s response time that the MWA can achieve
\citep{2015ApJ...814L..25K} we can easily repoint to catch any prompt
emission as long as the GW latency improves.  Overall we emphasize the
need to transmit the trigger and react, as soon as possible,
preferably well within 1\,min.

We have demonstrated that the MWA can respond quickly to GW transients
and cover a reasonable fraction of events with good sensitivity.  The strategies
outlined here are specifically applicable to the MWA, in that we have
made use of the MWA's location and primary beam pattern in assessing
the followup prospects.  They can  be adapted for other facilities,
but there other considerations may lead to different strategies.  For
instance, with a considerably smaller field-of-view but better
instantaneous sensitivity splitting into subarrays may be more
viable.  This will also evolve as new data and new models for prompt
emission become available.  Overall, we
believe that the MWA has a good combination of field-of-view,
sensitivity, and operational flexibility that enables this science:
the MWA has a much larger field-of-view compared to most pointed radio
telescopes \citep[e.g.,][]{2016MNRAS.459..121C,em2016}, but is more
sensitive than some all-sky facilities
\citep[e.g.,][]{2009IEEEP..97.1421E}.  With roughly 1 year of
sensitive GW observations we should be able to answer unambiguously
which if any of the models for prompt emission are real.

\begin{acknowledgements}
We thank an anonymous referee and editor for helpful comments.  
  DLK and SDC were supported by NSF grant AST-1412421.  Parts of this
research were conducted by the Australian Research Council Centre of
Excellence for All- sky Astrophysics (CAASTRO), through project number
CE110001020.  CMT is supported by an ARC Discovery Early Career
Researcher Project Grant, DE140100316.

\end{acknowledgements}

\bibliographystyle{apj} 

\end{document}